\documentclass[sigconf]{acmart}
\usepackage{graphicx}
\usepackage{subcaption}
\usepackage{enumitem}

\AtBeginDocument{%
  }

\setcopyright{none}
\copyrightyear{2024}
\acmYear{2024}
\acmDOI{X}

\acmConference[]{SDBD}{August 2024}{Barcelona,Spain }
\acmJournal{JDS}

\begin{document}


\settopmatter{printacmref=false}

\title{Benchmarking GNNs Using Lightning Network Data}

\author{Rainer Feichtinger}
\email{rainerfe@ethz.ch}
\affiliation{%
  \institution{ETH Zurich}
  \country{Switzerland}
}
\author{Florian Grötschla}
\email{fgroetschla@ethz.ch}
\affiliation{%
  \institution{ETH Zurich}
  \country{Switzerland}
}

\author{Lioba Heimbach}
\email{hlioba@ethz.ch}
\affiliation{%
  \institution{ETH Zurich}
  \country{Switzerland}
}

\author{Roger Wattenhofer}
\email{wattenhofer@ethz.ch}
\affiliation{%
  \institution{ETH Zurich}
  \country{Switzerland}
}

\renewcommand{\shortauthors}{Feichtinger et al.}
\acmArticleType{Research}
\authorsaddresses{}
\renewcommand\footnotetextcopyrightpermission[1]{} 

\acmContributions{BT and GKMT designed the study; LT, VB, and AP
  conducted the experiments, BR, HC, CP and JS analyzed the results,
  JPK developed analytical predictions, all authors participated in
  writing the manuscript.}

\begin{abstract}
    The Bitcoin Lightning Network is a layer 2 protocol designed to facilitate fast and inexpensive Bitcoin transactions. It operates by establishing channels between users, where Bitcoin is locked and transactions are conducted off-chain until the channels are closed, with only the initial and final transactions recorded on the blockchain. Routing transactions through intermediary nodes is crucial for users without direct channels, allowing these routing nodes to collect fees for their services. Nodes announce their channels to the network, forming a graph with channels as edges.
    In this paper, we analyze the graph structure of the Lightning Network and investigate the statistical relationships between node properties using machine learning, particularly Graph Neural Networks (GNNs). We formulate a series of tasks to explore these relationships and provide benchmarks for GNN architectures, demonstrating how topological and neighbor information enhances performance. Our evaluation of several models reveals the effectiveness of GNNs in these tasks and highlights the insights gained from their application.
\end{abstract}

\maketitle

\section{Introduction}

Bitcoin is a digital currency that relies on a distributed ledger known as the blockchain. This peer-to-peer network communicates via gossip messages and operates without a central intermediary. To ensure the security and integrity of transactions, the protocol uses Proof of Work as Sybil resistance. Miners, i.e., the block builders, are tasked with solving complex cryptographic puzzles to append new blocks to the blockchain.

Traditional centralized payment providers, e.g., Visa, can process tens of thousands of transactions per second. In contrast, Bitcoin is limited by its block size and average block time, which restricts its throughput to 3-10 transactions per second~\cite{Croman2016on}. Theoretically, these parameters can be adjusted to achieve higher throughput and lower latency. However, this would increase the bandwidth and hardware requirements for network participants and negatively impact the network's consistency, i.e., the guarantee that all honest parties output the same sequence of blocks. This trade-off presents a significant challenge for scaling Bitcoin to meet the demands of a global payment system.

The limited transaction throughput of Bitcoin leads to longer confirmation times and higher transaction fees, particularly during periods of high network activity. Thus, a scalable solution is needed to make Bitcoin practical for everyday use. Fundamentally, Bitcoin's scalability issue arises because every transaction is broadcast to the entire network, and each transaction's validity must be individually verified by all participants. This decentralized verification process is inherently inefficient for high-frequency transaction processing.

Layer-2 protocols, such as the Lightning Network, are designed to address this scalability problem. The key insight behind these protocols is that not all network participants need to be informed about and validate every transaction. The Lightning Network operates on the basis of payment channels. Two parties can open a payment channel by spending a certain amount of Bitcoin in a joint transaction called the funding transaction. Within this channel, they can conduct an unlimited number of transactions without immediately recording them on the blockchain. Only the funding transaction and the channel's closing transaction are recorded on the blockchain. These two on-chain transactions also enable the linking of payment channels to Bitcoin transactions and addresses.

The Lightning Network significantly reduces transaction fees and enables near-instant transactions since payments do not need to be confirmed by the Bitcoin blockchain. Another advantage is scalability; because most transactions occur off-chain, the Lightning Network can handle a substantially higher number of transactions per second. The complete details of a payment made through a channel are known only to the sender and receiver, offering greater privacy compared to on-chain transactions. 

Payment channels can be announced to the entire Lightning Network via gossip messages, allowing transactions between nodes that are not directly connected by a payment channel. The Lightning Network uses source routing based on the sender's local view of the network topology. However, routing is based on imperfect information because the current balances of channels (i.e., the distribution of capacity between two nodes) are not publicly available. This increases privacy but impacts routing efficiency. Moreover, not all payment channels are announced to the network; private channels exist, and little is known about their behavior and characteristics. Private channels add another layer of complexity to the network, making it challenging to obtain a complete picture of the network's topology and performance.

Machine learning techniques offer a promising approach to gaining deeper insights into the Lightning Network. For instance, a model capable of predicting channel balances could enhance routing efficiency. 

In this work, we present a benchmark based on Lightning Network data, demonstrating that the network's topological information can indeed be leveraged to predict certain properties. 

The Lightning Network can be interpreted as a graph, where the nodes of the Lightning Network are the vertices, and the payment channels are the edges of the graph. Various tasks can be defined on this graph, including regression and classification tasks at both the vertex and edge levels. As we demonstrate, Graph Neural Networks (GNNs) are particularly well-suited for solving these tasks. We show that GNNs can effectively utilize the topological information of the network to make predictions.

\section{Related Work}
The topology of the Lightning Network has been analyzed in several studies. For instance, Zabka et al.~\cite{zabka_centrality_2024}  and Seres et al.~\cite{seres_topological_2019} have examined its structural properties and centrality measures. Gossip messages within the Lightning Network were utilized by Zabka et al.~\cite{zabka_node_2021} to determine the implementation of Lightning nodes.

Romiti et al.~\cite{borisov_cross-layer_2021} demonstrated that the interconnection between the Lightning Network and the Bitcoin Network could be exploited to cluster Bitcoin addresses and link them to IP addresses. Additionally, Herrera-Joancomarti et al.~\cite{herrera-joancomarti_difficulty_2019} proposed a technique for discovering the balance of a Lightning channel. The application of machine learning techniques to predict channel balances was investigated by Vincent et al.~\cite{vincent_channel_2024}.
A dataset that links on-chain and off-chain data was presented by  Wang et al.~\cite{wang_etgraph_2023}. Specifically, data from Ethereum and Twitter were linked.

\section{Benchmark Introduction}

\begin{figure}
    \centering
    \includegraphics[width=1\linewidth]{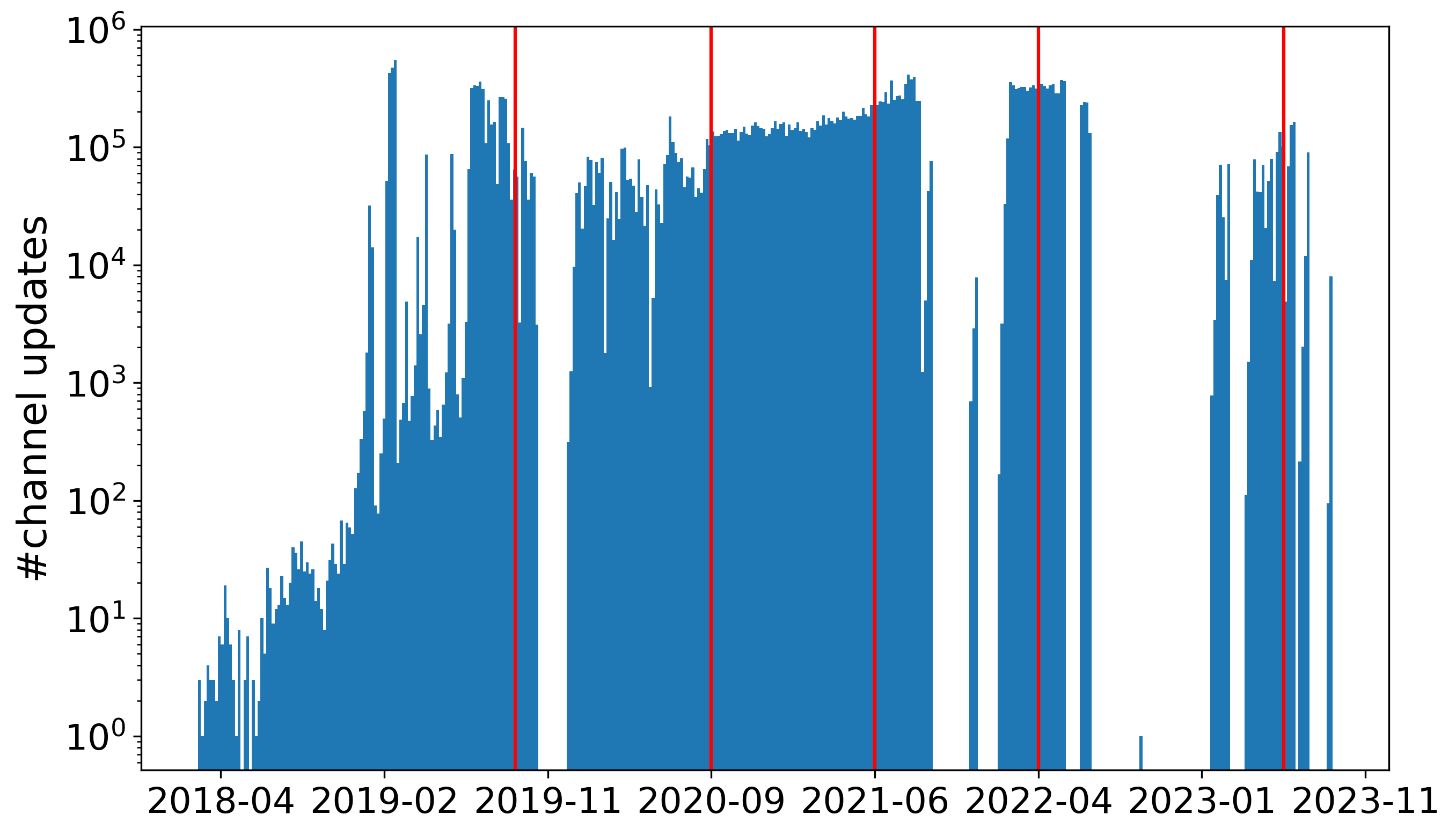}
    \caption{Channel Updates}
    \label{fig:channel_updates}
\end{figure}
\subsection{Gossip Messages}
Nodes in the Lightning Network use a gossip message protocol to exchange information about existing channels, facilitating pathfinding across various channels not directly operated by their own node. Through these gossip messages, information about the network's topology and details about individual nodes and channels are exchanged among the nodes. A \textbf{Channel Announcement} message informs the network about a new channel. It includes the signatures of both nodes involved, serving as proof that they agree on the creation of the channel within the public network. Additionally, it contains the Channel ID, which can be used to locate the on-chain funding transaction and thus determine the channel's capacity.

With \textbf{Node Announcement} messages, the nodes communicate at which address they can be reached. In addition, nodes indicate which features of the network the node supports and can also specify metadata such as an alias that can be freely selected by the node. 

A Channel Announcement alone is not sufficient for a channel to be used for routing in the Lightning Network. Only when both nodes involved in a channel publish details about the channel through channel updates does it become operational for routing. These \textbf{Channel Updates} communicate, among other things, the fees a node charges for routing through the channel. Additionally, they specify the minimum and maximum payment amounts that can be routed through the channel. The details of the channels can be regularly updated to reflect changes in the network topology or to adjust fees to remain attractive in path selection.

\subsection{Data Collection}

Since historical gossip messages are not stored on the blockchain, we have to rely on a source that has explicitly logged this data. By default, old gossip messages are also not stored by the nodes because they are replaced by newer messages and are not necessary for the functioning of the Lightning Network or the operation of a node. Our dataset is based on the Lightning Network Research Topology Dataset~\cite{lngossip}, which synchronized and stored gossip messages from the perspective of several nodes to achieve comprehensive coverage of the actual network.

Figure \ref{fig:channel_updates} shows the number of channel update messages over time. Noticeable are the gaps; during these periods, data logging did not function correctly. The red lines in the plot indicate the snapshots of the Lightning Network that we have chosen. These snapshots ensure that the gaps do not impact our dataset.

Additionally, we have extended the Lightning Network data. Using the IP addresses that nodes share in the Gossip Messages, we added location information for nodes that provided a valid IP address. For mapping IP addresses to locations, we used \texttt{ipinfo.io}. Moreover, we linked the Lightning Network data with blockchain data. For each channel, we added the capacity by referencing the funding transaction on the Bitcoin blockchain.

\subsection{Dataset}

\begin{tabular}{llllll}
\toprule
 & 2019-10 & 2020-09 & 2021-06 & 2022-04 & 2023-06 \\
\midrule
nodes & 4,740 & 5,990 & 10,835 & 18,746 & 15,287 \\
edges & 51,414 & 52,187 & 81,389 & 151,092 & 116,067 \\
avg. degree & 10.85 & 8.71 & 7.51 & 8.06 & 7.59 \\
weakly cc & 4 & 9 & 35 & 76 & 29 \\
diameter & 7 & 8 & 9 & 9 & 9 \\
\bottomrule
\label{table:dataset}
\end{tabular}

Table \ref{table:dataset} shows an overview and important properties of the graphs of our dataset at the respective points in time. Whereby for all snapshots over 99\% of the nodes are in the largest weakly connected component. The fact that there are several connected components does not have a significant influence on the performance of the models, because the other connected components are negligible compared to the largest connected component.

In this section, we aim to provide a detailed description of all the features used in the benchmark: 
\begin{itemize}
    \item \textbf{CLTV Expiry Delta}
    the CLTV (Check Lock Time Verify) expiry delta is the number of blocks a node can wait before the node risks losing BTC in the event of a delayed transaction on the Lightning Network.
    \item \textbf{Minimum HTLC} is the minimum value in millisatoshis per HTCL (Hashed Timelock Contract)  that can be routed via this channel.
    \item \textbf{Maximum HTLC} is the maximum value in millisatoshis per HTCL that can be routed via this channel.
    \item \textbf{Base Fee} is the fixed amount in millisatoshis charged by a node for forwarding a payment, regardless of the payment size.
    \item \textbf{Proportional Fee} is the amount in millisatoshis charged by a node for forwarding a payment, per transferred Satoshi.
\item \textbf{RGB Color} This value can be freely selected by each node. The value is not relevant for the protocol itself but can be used to visualize the network.  Since different implementations of the Lightning Network software have varying default RGB color values, these values can offer insights into the specific implementation a node might be using.
    \item \textbf{Country} The country of a node can be identified if the node provides an IP address via gossip messages. 
    \item \textbf{Capacity} in mBTC is the amount with which a channel is created on-chain through a funding transaction. This value, therefore, indicates the size of the channel but does not indicate the current distribution of capacity between the two nodes involved in a channel.
\end{itemize}

In summary, CLTV Expiry Delta, Minimum HTLC, Maximum HTLC, Base Fee, and Proportional Fee are channel values that can be extracted from the Lightning gossip messages. The capacity is also a value that relates to a channel but is extracted from the Bitcoin blockchain. The country and the RGB color value are properties of the nodes and can be extracted from gossip messages or, in the case of the country, derived from the IP address using additional data sources. There are also additional features that can be extracted from the gossip messages. However, we have deliberately limited our focus to these features because they are particularly informative and facilitate a straightforward interpretation of the tasks, even without detailed knowledge of the exact mechanics of the Lightning Network.

\section{Benchmark Evaluation}
We evaluated different models for each task, including graph convolution network \cite{kipf2017semisupervised}, graph attention network \cite{veličković2018graph}, graph isomorphism network \cite{xu2019powerful}, the modified graph isomorphism network capable of incorporating edge features \cite{hu2020strategies}, and a variant where edge features are concatenated with the embeddings of the involved nodes only after the final message passing layer. Additionally, we included the GraphSAGE network 
\cite{hamilton2018inductive} and residual gated graph ConvNets\cite{bresson2018residual}. Furthermore, we evaluated Graph Transformer models from the GRAPHGPS \cite{rampášek2023recipe} framework.

As a baseline, we used an MLP model, and for tasks utilizing mean absolute error (MAE) as a criterion, we also employed the median as a naive predictor. The evaluation was conducted on a fixed 10/20/70 validation/test/train split, with three runs using different seeds. Due to the number of different models, tasks, and snapshots of the data, we did not perform an exhaustive search for hyperparameters. Instead, we conducted a controlled random search using GraphGym \cite{you2021design} and the recommended design dimensions. For all tasks, the performance of the GNN models is better than the baseline.
\subsection{Tor Classification}
\begin{figure}[h]
    \centering
    \includegraphics[width=0.7\linewidth]{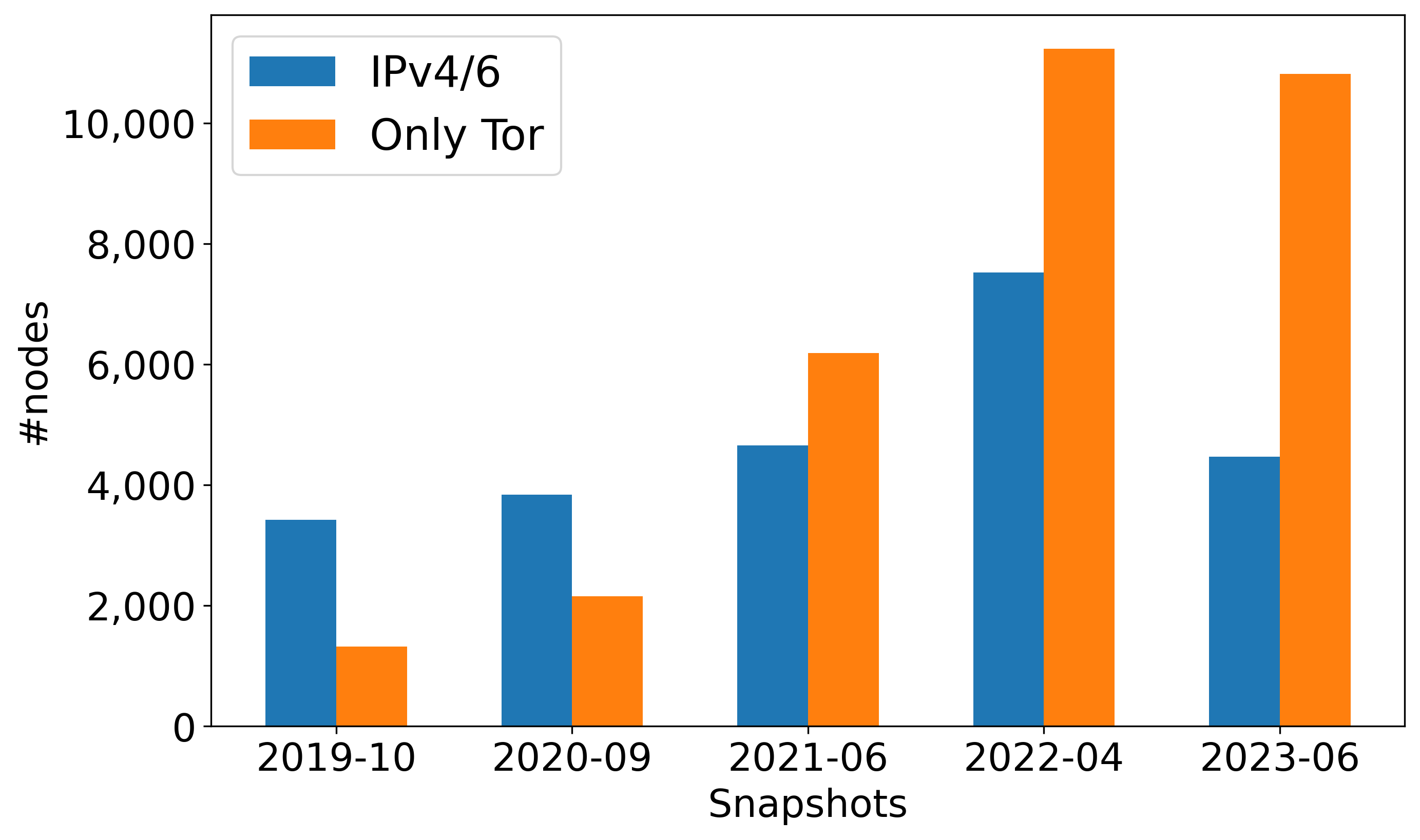}
    \caption{Node Address}
    \label{fig:ip-vs-tor-address}
\end{figure}

The Lightning Network supports both IP addresses and Tor addresses. Figure \ref{fig:ip-vs-tor-address} illustrates this, with the number of nodes using an IP address depicted in blue and the number of nodes exclusively using a Tor address shown in orange. The data reveal a notable increase over time in the number of nodes that exclusively use a Tor address in our snapshots.
The hypothesis for the task is that nodes that only specify a Tor address are particularly interested in privacy and can be distinguished from nodes that specify an IP address based on their behavior and topology in the Lightning Network.

\begin{figure}
    \centering
    \includegraphics[width=0.7\linewidth]{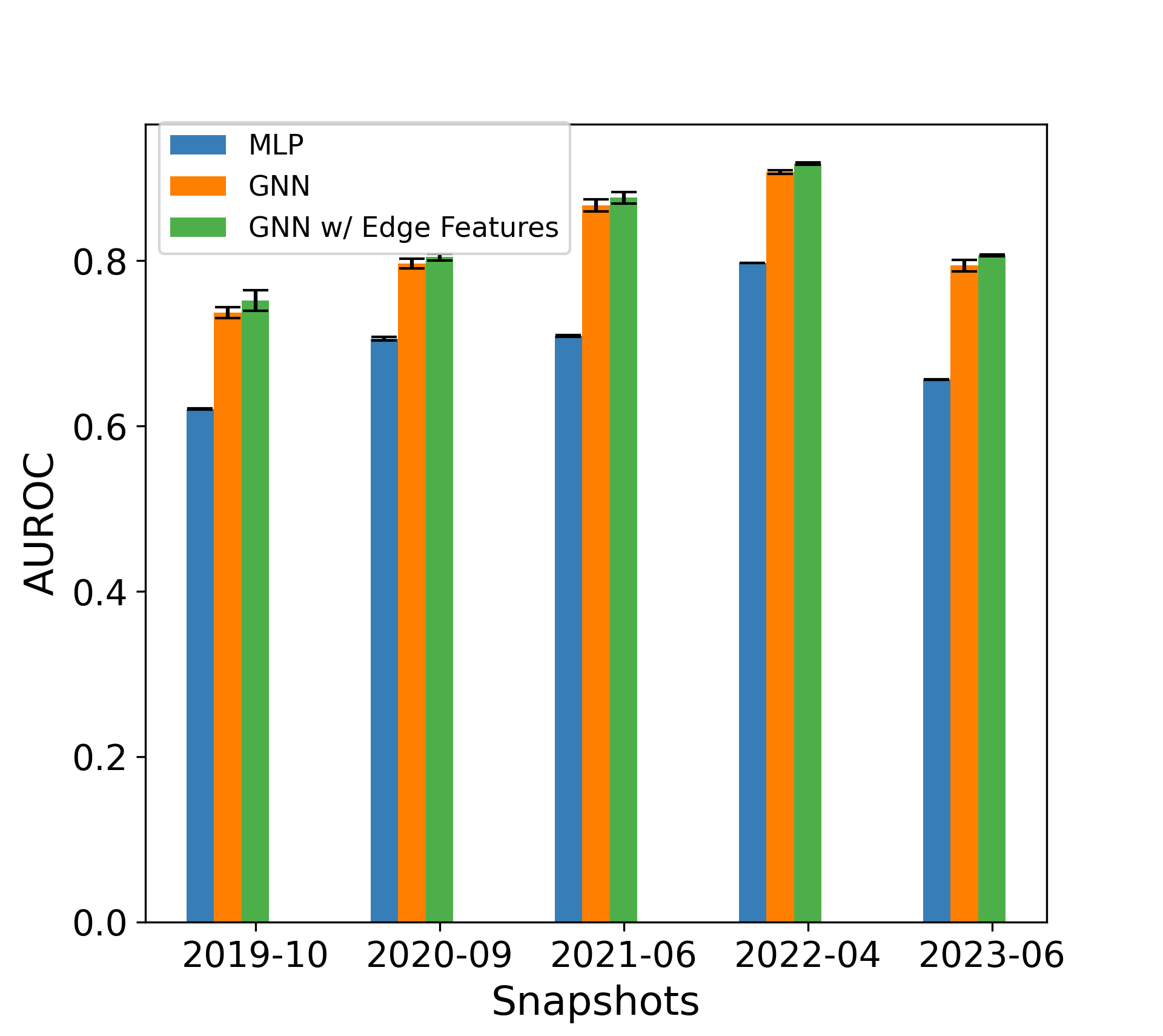}
    \caption{Tor Classification}
    \label{fig:torClassification}
\end{figure}

Figure \ref{fig:torClassification} shows the performance of the models for this task using the AUROC score as the criterion. We observe that the models can indeed distinguish nodes that exclusively provide a Tor address from those that provide an IP address. GNNs that utilize edge features performed slightly better than GNNs without edge features across all snapshots. Among the models with edge features, GatedGCNConv was the best model for all snapshots, while among the models without edge features, GINConv consistently had the best performance except for the snapshot from June 2021, where GATConv performed slightly better

\subsection{Capacity Regression}
 \label{subsection:capacity_Regression}

\begin{figure}[h]
    \centering
    \begin{subfigure}[b]{0.49\linewidth}
        \centering
        \includegraphics[width=\linewidth]{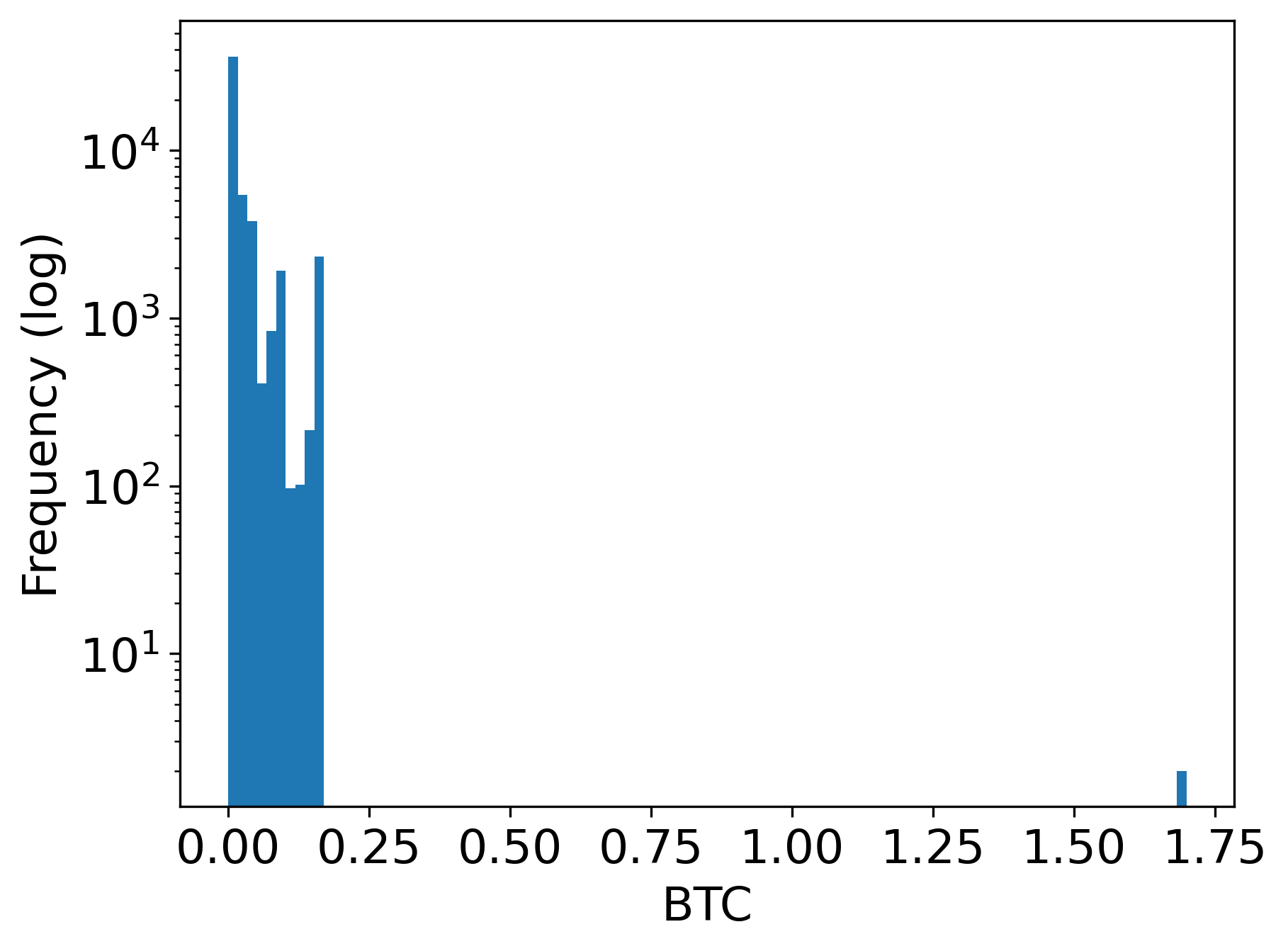}
        \label{fig:capacity_dist_1570000000}
    \end{subfigure}
    \hfill
    \begin{subfigure}[b]{0.48\linewidth}
        \centering
        \includegraphics[width=\linewidth]{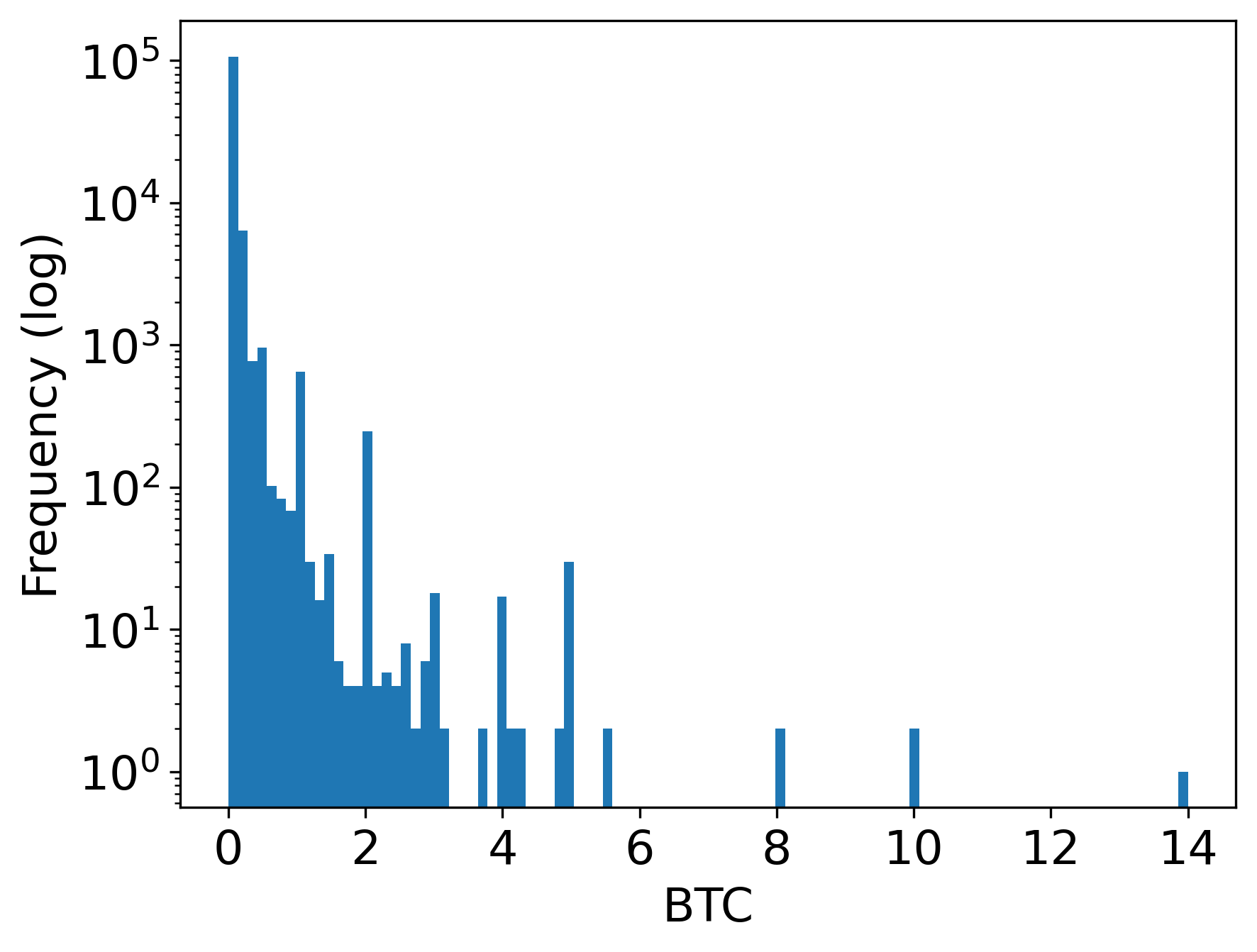}
        \label{fig:capacity_dist_1687500000}
    \end{subfigure}
    \caption{Capacity Distribution}
    \label{fig:capacity_distribution}
\end{figure}

Figure \ref{fig:capacity_distribution} illustrates the distribution of channel capacities at our earliest (left) and latest (right) snapshots. The average capacity was 0.0231 BTC at the earliest snapshot and increased to 0.0609 BTC at the latest snapshot. This indicates that the capacity of channels has grown over time. However, the majority of channels still have a capacity of less than 0.1 BTC, with 94.62\% at the earliest snapshot and 89.97\% at the latest snapshot.

\begin{figure}
    \centering
    \includegraphics[width=0.7\linewidth]{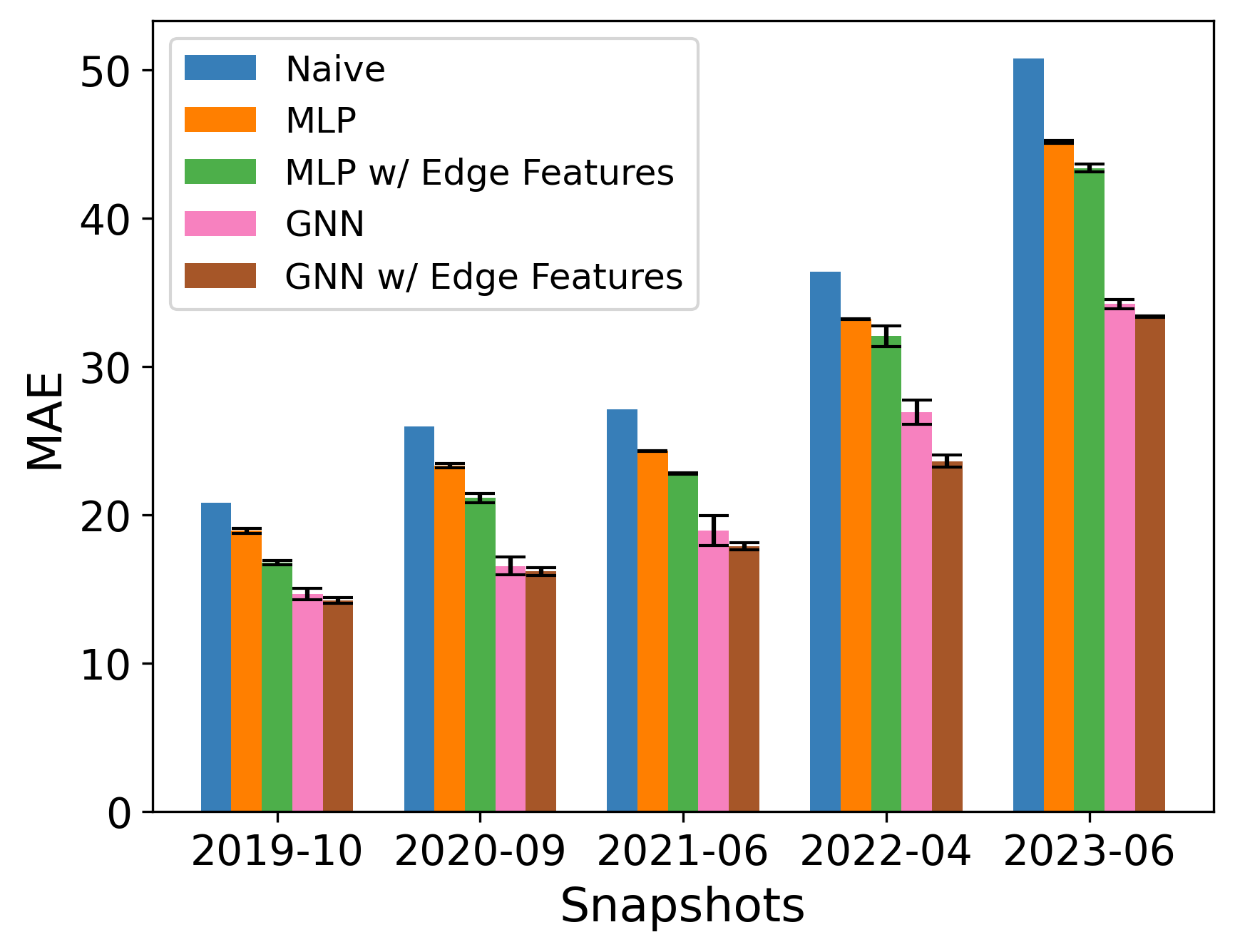}
    \caption{Capacity Regression}
    \label{fig:capacity}
\end{figure}

Figure \ref{fig:capacity} shows the MAE for various models across different snapshots. The capacity and consequently the MAE are given in mBTC (1 BTC = 1000 mBTC). We observe that the MAE increased over time, which can be explained by the rise in the average capacity and the expanding range of capacity values. Furthermore, we note that the GNN models utilizing edge features perform better across all snapshots compared to those without edge features. Except for the earliest snapshot, the best-performing GNN model without edge features was consistently GINConv, while for the first snapshot, GraphSAGE performed slightly better. The best performance for GNN models using edge features was predominantly delivered by a GPS model with GatedGCN and Laplacian as positional encodings. However, for the first snapshot, the GPS model with random-walk structural encoding performed better, and for the second snapshot, GINConv with edge features concatenated after the last message passing layer was superior.

\subsection{Base Fee Regression}
\label{subsection:base_fee}
\begin{figure}[h]
    \centering
    \includegraphics[width=0.7\linewidth]{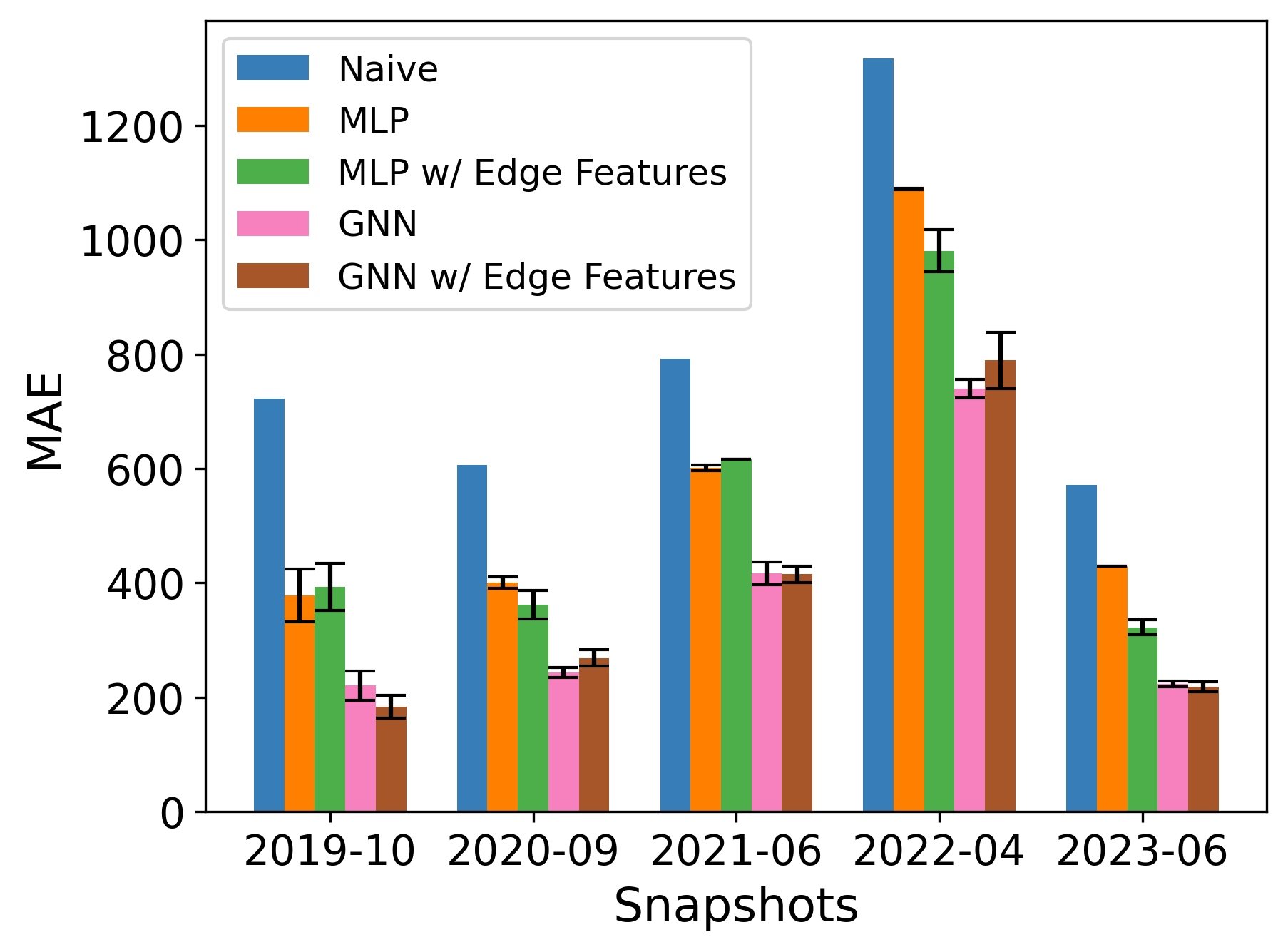}
    \caption{Base Fee Regression}
    \label{fig:LightningFeeBase}
\end{figure}
The base fee influences routing within the Lightning Network. Our observations indicate that extremely high base fee values occasionally occur. The network’s specifications allow for a base fee value of type uint32. In some cases, even the maximum value of uint32 was used, which would make transactions through these channels significantly more expensive than regular Bitcoin transactions. We exclude channels with extremely high base fees as they are irrelevant for routing. However, over 99\% of channels are retained in all snapshots.

Figure \ref{fig:LightningFeeBase} displays the MAE in millisatoshis for the prediction of the base fee. Across all snapshots, the GINConv model without edge features performed the best. Among the models with edge features, a GINConv model where edge features were added after the last message-passing layer performed the best for the first four snapshots. In the final snapshot, a GatedGNCConv model had the best performance.

\subsection{Proportional Fee Regression}

Similar to the base fee, the proportional fee also influences routing within the Lightning Network. We have observed issues with high values here as well and have filtered out these extreme cases. Nevertheless, over 99\% of channels are still included in this scenario.

\begin{figure}[h]
    \centering
    \includegraphics[width=0.7\linewidth]{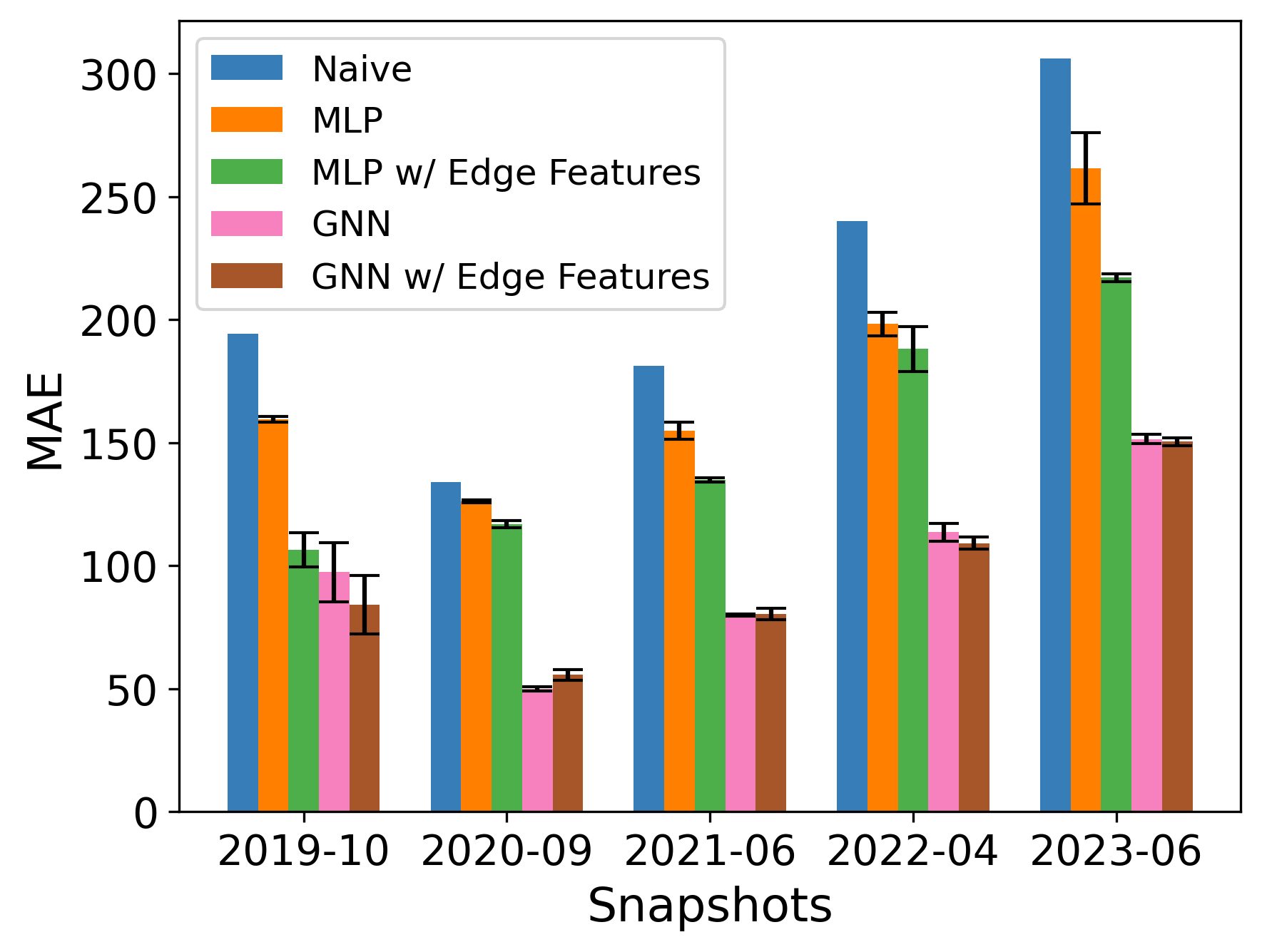}
    \caption{Proportional Fee Regression}
    \label{fig:LightningFeeProportional}
\end{figure}

Figure \ref{fig:LightningFeeProportional} shows the performance of the models measured in MAE in millisatoshis. Similar to the prediction of the base fee, the performance of GINConv models is particularly good here. Only in the first snapshot was a GraphSAGE model slightly better. Among the models with edge features, a GINConv model with edge features was the best in all snapshots.

\subsection{Maximum HTLC Regression}
The HTLC maximum value can be freely chosen by nodes for each channel but must not exceed the channel's capacity. In our dataset, it is observed that the HTLC maximum value is predominantly very close to the channel's capacity. Therefore, the distribution of HTLC maximum values is quite similar to the distribution of capacity values. Consequently, the model performance is similar to that in (\ref{subsection:capacity_Regression}) capacity regression.

\begin{figure}[h]
    \centering
    \includegraphics[width=0.7\linewidth]{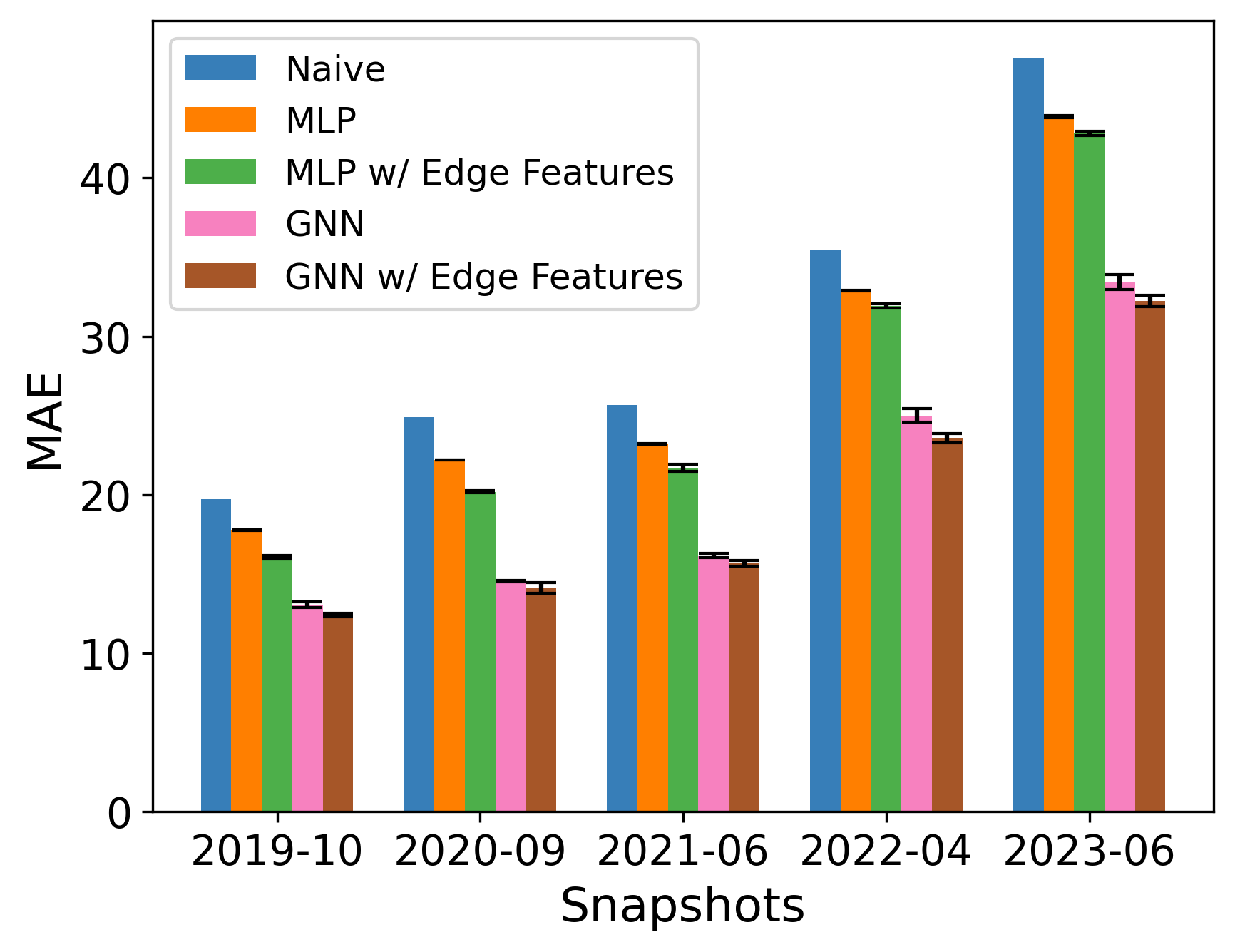}
    \caption{Maximum HTLC Regression}
    \label{fig:LightningHtlcMaximum}
\end{figure}

Figure \ref{fig:LightningHtlcMaximum} shows the MAE for different snapshots and models. For better comparability with section \ref{subsection:capacity_Regression}, the HTLC maximum values and thus the MAE are given in mBTC. For the first three snapshots, GINconv was the best-performing model without edge features, whereas for the last two snapshots, GraphSAGE performed slightly better. Among the models utilizing edge features, the GPS model with GatedGCNConv and Laplacian encoding showed the best performance for the last two snapshots. A GINconv model, which adds edge features after the message-passing layer, had the best performance for the earliest snapshot, while GatedGCNconv performed best for the snapshots in 2020-09 and 2021-06.

\subsection{Link Prediction}
In this task, the goal is to predict whether a channel exists between two nodes. For our ground truth, we limit ourselves to public channels. However, good models could potentially be used in the future to find candidates for private channels.

\begin{figure}[h]
    \centering
    \includegraphics[width=0.7\linewidth]{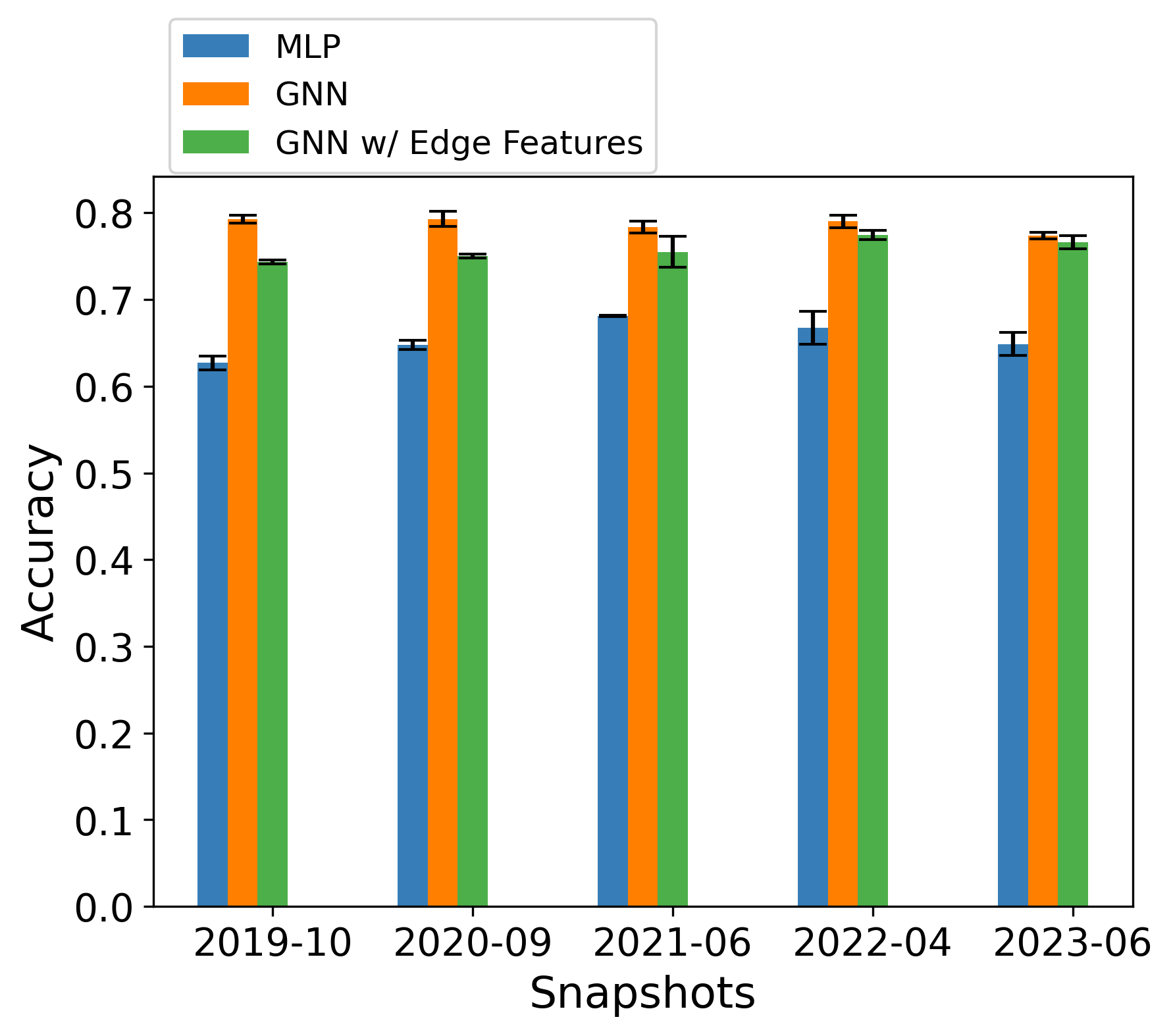}
    \caption{Link Prediction}
    \label{fig:LinkPrediction}
\end{figure}

Figure \ref{fig:LinkPrediction} shows the accuracy achieved by the models in this task. Notably, the models with edge features perform worse than those without.

The highest accuracy for models with edge features was achieved by a GatedGCNconv model for all snapshots. Without edge features, the performance was best for the first three snapshots with a GraphSAGE model and for the remaining snapshots with a GINConv model.

\newpage
\section{Conclusion}

In conclusion, we have linked Lightning Network data with Bitcoin data and demonstrated that this combined dataset is well-suited as a benchmark for Graph Neural Networks. We defined and evaluated various tasks, including regression and classification at both vertex and edge levels. Our findings indicate that GNNs can effectively leverage topological and neighborhood information to enhance performance in these tasks. The diversity of tasks and the real-world data from a payment network make this benchmark particularly well-suited for testing and comparing different GNN models. Moreover, robust models developed through this benchmark could provide deeper insights into the dynamics of the Lightning Network in the future.

\bibliographystyle{ACM-Reference-Format}
\bibliography{references}

\newpage

\onecolumn
\appendix
\pagenumbering{alph}
\section{Appendix}

The following tables provide the results for the tasks described in the paper. Each entry includes the model’s performance and the standard deviation.

\begin{table}[h]
\centering
\begin{tabular}{llllll}
\toprule
 & 2019-10 & 2020-09 & 2021-06 & 2022-04 & 2023-06 \\
 \hline
Model  & MAE & MAE & MAE & MAE & MAE \\
\midrule
GPS + GatedGCNConv w/ LapPE & 14.378$\pm$0.100 & 16.750$\pm$0.038 & 17.880$\pm$0.220 & 23.605$\pm$0.407 & 33.349$\pm$0.055 \\
GPS + GatedGCNConv w/ RSWE & 14.230$\pm$0.169 & 16.875$\pm$0.045 & 18.196$\pm$0.309 & 24.118$\pm$0.315 & 33.811$\pm$0.575 \\
GPS + GINE w/ RSWE & 14.551$\pm$0.114 & 16.808$\pm$0.118 & 18.143$\pm$0.205 & 24.160$\pm$0.147 & 34.500$\pm$0.172 \\
GATConv & 15.890$\pm$1.063 & 18.763$\pm$0.944 & 20.803$\pm$0.653 & 27.781$\pm$0.798 & 36.009$\pm$0.246 \\
GatedGCNConv & 17.489$\pm$0.525 & 20.708$\pm$0.581 & 22.174$\pm$0.202 & 29.270$\pm$0.092 & 38.671$\pm$0.427 \\
GCNConv & 15.332$\pm$0.245 & 18.318$\pm$0.479 & 20.185$\pm$0.794 & 28.153$\pm$0.278 & 36.691$\pm$0.823 \\
GINConv & 14.789$\pm$0.368 & 16.558$\pm$0.620 & 18.920$\pm$1.007 & 26.918$\pm$0.822 & 34.207$\pm$0.307 \\
GINConv w/ Edge Features & 14.339$\pm$0.132 & 16.178$\pm$0.263 & 18.882$\pm$0.572 & 26.022$\pm$0.685 & 34.592$\pm$1.231 \\
GINEConv & 19.973$\pm$0.036 & 24.284$\pm$0.927 & 23.553$\pm$0.211 & 30.819$\pm$0.411 & 43.786$\pm$0.484 \\
MLP & 18.921$\pm$0.162 & 23.324$\pm$0.150 & 24.297$\pm$0.038 & 33.202$\pm$0.017 & 45.117$\pm$0.085 \\
MLP w/ Edge Features & 16.763$\pm$0.135 & 21.142$\pm$0.306 & 22.782$\pm$0.032 & 32.038$\pm$0.697 & 43.349$\pm$0.265 \\
GraphSAGE & 14.659$\pm$0.397 & 17.061$\pm$0.655 & 19.062$\pm$0.450 & 27.130$\pm$0.230 & 35.564$\pm$0.872 \\
Naive & 20.8140 & 25.9383 & 27.0931 & 36.3631 & 50.7388 \\
\bottomrule
\end{tabular}
\caption{Capacity Regression}
\label{tab:LightningCapacity}
\end{table}

\begin{table}[h]
\centering
\begin{tabular}{llllll}
\toprule
 & 2019-10 & 2020-09 & 2021-06 & 2022-04 & 2023-06 \\
 \hline
Model  & MAE & MAE & MAE & MAE & MAE \\
\midrule
GPS + GatedGCNConv w/ LapPE & 13.012$\pm$0.106 & 15.642$\pm$0.237 & 16.166$\pm$0.159 & 23.571$\pm$0.302 & 32.220$\pm$0.366 \\
GPS + GatedGCNConv w/ RSWE & 13.050$\pm$0.047 & 15.948$\pm$0.119 & 16.563$\pm$0.190 & 24.607$\pm$0.386 & 33.066$\pm$0.231 \\
GPS + GINE w/ RSWE & 13.377$\pm$0.277 & 16.036$\pm$0.223 & 16.888$\pm$0.223 & 24.579$\pm$0.343 & 33.929$\pm$0.369 \\
GATConv & 15.115$\pm$0.270 & 17.873$\pm$0.150 & 18.717$\pm$0.227 & 28.264$\pm$0.265 & 38.941$\pm$0.847 \\
GatedGCNConv & 12.670$\pm$0.071 & 14.120$\pm$0.333 & 15.669$\pm$0.183 & 24.112$\pm$0.442 & 32.522$\pm$0.404 \\
GCNConv & 14.281$\pm$0.180 & 16.650$\pm$0.320 & 18.617$\pm$0.365 & 27.636$\pm$0.060 & 36.915$\pm$0.536 \\
GINConv & 13.066$\pm$0.187 & 14.558$\pm$0.039 & 16.152$\pm$0.127 & 25.151$\pm$0.315 & 33.645$\pm$0.149 \\
GINConv w/ Edge Features & 12.405$\pm$0.119 & 14.558$\pm$0.479 & 16.134$\pm$0.576 & 25.349$\pm$0.193 & 33.845$\pm$0.717 \\
GINEConv & 19.027$\pm$0.297 & 23.897$\pm$0.157 & 24.004$\pm$0.469 & 31.932$\pm$0.414 & 44.552$\pm$0.342 \\
MLP & 17.770$\pm$0.018 & 22.192$\pm$0.020 & 23.189$\pm$0.022 & 32.868$\pm$0.019 & 43.868$\pm$0.076 \\
MLP w/ Edge Features & 16.079$\pm$0.086 & 20.175$\pm$0.071 & 21.690$\pm$0.234 & 31.918$\pm$0.141 & 42.784$\pm$0.136 \\
GraphSAGE & 13.423$\pm$0.163 & 14.584$\pm$0.321 & 16.178$\pm$0.241 & 24.993$\pm$0.422 & 33.438$\pm$0.471 \\
Naive & 19.7165 & 24.8738 & 25.6508 & 35.4004 & 47.5202 \\
\bottomrule
\end{tabular}
\caption{Maximum HTLC Regression}
\label{tab:LightningHtlcMaximum}
\end{table}

\begin{table}[h]
\centering
\begin{tabular}{llllll}
\toprule
 & 2019-10 & 2020-09 & 2021-06 & 2022-04 & 2023-06 \\
 \hline
Model  & Accuracy & Accuracy & Accuracy & Accuracy & Accuracy \\
\midrule
GPS + GatedGCNConv w/ LapPE & 0.723$\pm$0.007 & 0.748$\pm$0.019 & 0.713$\pm$0.002 & 0.758$\pm$0.012 & 0.704$\pm$0.008 \\
GPS + GatedGCNConv w/ RSWE & 0.716$\pm$0.013 & 0.734$\pm$0.003 & 0.739$\pm$0.013 & 0.765$\pm$0.003 & 0.736$\pm$0.001 \\
GPS + GINE w/ RSWE & 0.733$\pm$0.012 & 0.732$\pm$0.007 & 0.738$\pm$0.008 & 0.734$\pm$0.003 & 0.734$\pm$0.005 \\
GATConv & 0.641$\pm$0.050 & 0.642$\pm$0.042 & 0.667$\pm$0.039 & 0.730$\pm$0.026 & 0.651$\pm$0.021 \\
GatedGCNConv & 0.743$\pm$0.002 & 0.750$\pm$0.002 & 0.755$\pm$0.018 & 0.774$\pm$0.005 & 0.766$\pm$0.008 \\
GCNConv & 0.764$\pm$0.009 & 0.764$\pm$0.009 & 0.784$\pm$0.004 & 0.795$\pm$0.007 & 0.776$\pm$0.003 \\
GINConv & 0.761$\pm$0.001 & 0.775$\pm$0.005 & 0.777$\pm$0.003 & 0.790$\pm$0.007 & 0.774$\pm$0.004 \\
GINEConv & 0.700$\pm$0.024 & 0.687$\pm$0.011 & 0.727$\pm$0.005 & 0.755$\pm$0.007 & 0.737$\pm$0.017 \\
MLP & 0.627$\pm$0.008 & 0.648$\pm$0.005 & 0.681$\pm$0.001 & 0.668$\pm$0.019 & 0.649$\pm$0.013 \\
GraphSAGE & 0.793$\pm$0.005 & 0.793$\pm$0.009 & 0.783$\pm$0.007 & 0.774$\pm$0.008 & 0.771$\pm$0.007 \\
\bottomrule
\end{tabular}
\caption{Link Prediction}
\label{tab:LinkPrediction}
\end{table}

\begin{table}[h]
\centering
\begin{tabular}{llllll}
\toprule
 & 2019-10 & 2020-09 & 2021-06 & 2022-04 & 2023-06 \\
 \hline
Model  & AUROC & AUROC & AUROC & AUROC & AUROC \\
\midrule
GPS + GatedGCNConv w/ LapPE & 0.632$\pm$0.013 & 0.721$\pm$0.005 & 0.788$\pm$0.020 & 0.864$\pm$0.003 & 0.725$\pm$0.013 \\
GPS + GatedGCNConv w/ RSWE & 0.700$\pm$0.010 & 0.741$\pm$0.002 & 0.824$\pm$0.015 & 0.881$\pm$0.011 & 0.715$\pm$0.010 \\
GPS + GINE w/ RSWE & 0.705$\pm$0.014 & 0.761$\pm$0.019 & 0.840$\pm$0.003 & 0.896$\pm$0.002 & 0.757$\pm$0.019 \\
GATConv & 0.723$\pm$0.008 & 0.795$\pm$0.008 & 0.867$\pm$0.007 & 0.904$\pm$0.006 & 0.782$\pm$0.033 \\
GatedGCNConv & 0.752$\pm$0.012 & 0.804$\pm$0.004 & 0.876$\pm$0.007 & 0.918$\pm$0.001 & 0.807$\pm$0.001 \\
GCNConv & 0.728$\pm$0.005 & 0.767$\pm$0.007 & 0.844$\pm$0.006 & 0.897$\pm$0.002 & 0.794$\pm$0.004 \\
GINConv & 0.737$\pm$0.007 & 0.797$\pm$0.006 & 0.855$\pm$0.002 & 0.907$\pm$0.002 & 0.794$\pm$0.007 \\
GINEConv & 0.717$\pm$0.017 & 0.764$\pm$0.011 & 0.821$\pm$0.006 & 0.857$\pm$0.009 & 0.772$\pm$0.004 \\
MLP & 0.621$\pm$0.001 & 0.706$\pm$0.002 & 0.709$\pm$0.001 & 0.798$\pm$0.000 & 0.656$\pm$0.000 \\
GraphSAGE & 0.714$\pm$0.021 & 0.786$\pm$0.010 & 0.858$\pm$0.004 & 0.897$\pm$0.004 & 0.782$\pm$0.003 \\
\bottomrule
\end{tabular}
\caption{Tor Classification}
\label{tab:TorPrediction}
\end{table}

\begin{table}[h]
\centering
\begin{tabular}{llllll}
\toprule
 & 2019-10 & 2020-09 & 2021-06 & 2022-04 & 2023-06 \\
 \hline
Model  & MAE & MAE & MAE & MAE & MAE \\
\midrule
GPS + GatedGCNConv w/ LapPE & 537.199$\pm$49.986 & 423.320$\pm$15.841 & 620.814$\pm$25.198 & 789.047$\pm$49.129 & 325.947$\pm$3.285 \\
GPS + GatedGCNConv w/ RSWE & 584.988$\pm$16.159 & 413.693$\pm$19.733 & 651.881$\pm$13.142 & 790.772$\pm$49.424 & 326.789$\pm$5.033 \\
GPS + GINE w/ RSWE & 566.279$\pm$3.424 & 452.159$\pm$21.018 & 646.356$\pm$2.688 & 978.579$\pm$5.791 & 346.237$\pm$6.307 \\
GATConv & 475.971$\pm$9.179 & 366.554$\pm$14.257 & 621.617$\pm$11.276 & 991.366$\pm$60.510 & 313.897$\pm$18.376 \\
GatedGCNConv & 412.994$\pm$30.591 & 306.872$\pm$14.778 & 521.986$\pm$11.463 & 810.773$\pm$9.201 & 218.660$\pm$8.609 \\
GCNConv & 443.883$\pm$3.210 & 350.189$\pm$2.765 & 589.511$\pm$1.946 & 898.937$\pm$7.084 & 292.707$\pm$13.659 \\
GINConv & 220.686$\pm$25.346 & 243.477$\pm$8.686 & 416.764$\pm$20.211 & 739.607$\pm$15.774 & 223.311$\pm$4.583 \\
GINConv w/ Edge Features & 182.940$\pm$19.996 & 268.526$\pm$14.195 & 414.922$\pm$13.856 & 790.090$\pm$40.303 & 232.136$\pm$2.961 \\
GINEConv & 540.510$\pm$7.611 & 473.277$\pm$0.024 & 792.366$\pm$0.001 & 1155.582$\pm$21.106 & 504.282$\pm$1.290 \\
MLP & 378.040$\pm$45.908 & 400.791$\pm$10.217 & 601.548$\pm$5.009 & 1089.004$\pm$1.040 & 429.026$\pm$0.209 \\
MLP w/ Edge Features & 392.683$\pm$40.950 & 361.775$\pm$24.395 & 616.062$\pm$0.351 & 980.669$\pm$36.470 & 322.344$\pm$12.817 \\
GraphSAGE & 264.330$\pm$23.160 & 295.983$\pm$6.726 & 438.151$\pm$12.335 & 843.571$\pm$5.587 & 262.662$\pm$5.161 \\
Naive & 722.3765 & 605.6976 & 792.3601 & 1316.7023 & 571.5372 \\
\bottomrule
\end{tabular}
\caption{Base Fee Regression}
\label{tab:LightningFeeBase}
\end{table}

\begin{table}[h]
\centering
\begin{tabular}{llllll}
\toprule
 & 2019-10 & 2020-09 & 2021-06 & 2022-04 & 2023-06 \\
 \hline
Model  & MAE & MAE & MAE & MAE & MAE \\
\midrule
GPS + GatedGCNConv w/ LapPE & 278.849$\pm$102.170 & 133.641$\pm$0.954 & 162.458$\pm$0.973 & 214.890$\pm$0.478 & 264.229$\pm$0.286 \\
GPS + GatedGCNConv w/ RSWE & 302.676$\pm$85.778 & 135.228$\pm$6.113 & 160.678$\pm$0.802 & 215.577$\pm$0.877 & 263.368$\pm$0.962 \\
GPS + GINE w/ RSWE & 826.913$\pm$345.514 & 136.108$\pm$2.317 & 161.940$\pm$1.751 & 213.538$\pm$1.347 & 261.932$\pm$0.426 \\
GATConv & 140.708$\pm$14.914 & 131.415$\pm$2.625 & 124.310$\pm$7.679 & 190.759$\pm$4.481 & 238.401$\pm$8.236 \\
GatedGCNConv & 98.690$\pm$6.548 & 76.813$\pm$1.609 & 81.613$\pm$0.957 & 117.268$\pm$2.085 & 152.864$\pm$3.057 \\
GCNConv & 101.823$\pm$5.475 & 126.858$\pm$2.030 & 181.321$\pm$0.015 & 226.995$\pm$1.786 & 222.385$\pm$3.118 \\
GINConv & 100.084$\pm$1.270 & 49.865$\pm$0.846 & 79.917$\pm$0.567 & 113.666$\pm$3.647 & 151.524$\pm$2.024 \\
GINConv w/ Edge Features & 84.233$\pm$11.907 & 55.574$\pm$2.125 & 80.354$\pm$2.317 & 109.152$\pm$2.366 & 150.404$\pm$1.507 \\
GINEConv & 194.182$\pm$0.000 & 134.098$\pm$0.000 & 180.553$\pm$0.225 & 235.490$\pm$0.862 & 300.852$\pm$1.435 \\
MLP & 159.498$\pm$1.295 & 126.164$\pm$0.655 & 154.766$\pm$3.522 & 198.318$\pm$4.784 & 261.541$\pm$14.565 \\
MLP w/ Edge Features & 106.416$\pm$6.856 & 116.920$\pm$1.513 & 134.780$\pm$0.914 & 188.119$\pm$9.057 & 217.067$\pm$1.568 \\
GraphSAGE & 97.343$\pm$11.966 & 94.297$\pm$9.293 & 85.964$\pm$2.142 & 135.467$\pm$4.117 & 175.291$\pm$7.586 \\
Naive & 194.1822 & 134.0981 & 181.3260 & 240.1717 & 306.1413 \\
\bottomrule
\end{tabular}
\caption{Proportional Fee Regression}
\label{tab:LightningFeeProportional}
\end{table}

\end{document}